\documentclass[12pt]{article}
\usepackage{amsmath}       
\usepackage{amssymb}       
\usepackage{a4wide}        
\usepackage{float}         
\usepackage{graphicx}      
\usepackage{bbm}           
\usepackage[small,bf]{caption}       
\usepackage{varioref}
\usepackage{dcolumn}        
\usepackage{cite}           
\usepackage{color}
\pdfoutput=1
\definecolor{red}{rgb}{1,0,0}
\def\red{}

\author{Peter Brommer\thanks{e-mail:
    \texttt{p.brommer@itap.physik.uni-stuttgart.de}} and 
Franz G\"ahler\\
  Institut f\"ur Theoretische und Angewandte Physik\\
  Universit\"at Stuttgart} \date{\today}

\title{Effective potentials for quasicrystals from ab-initio data}
\begin{document}

\maketitle

\begin{abstract}
Classical effective potentials are indispensable for any large-scale
atomistic simulations, and the relevance of simulation results
crucially depends on the quality of the potentials used. For complex 
alloys like quasicrystals, however, realistic effective potentials are 
practically inexistent. We report here on our efforts to develop 
effective potentials especially for quasicrystalline alloy systems.
We use the so-called force matching method, in which the potential
parameters are adapted so as to optimally reproduce the forces
and energies in a set of suitably chosen reference configurations.
These reference data are calculated with ab-initio methods.
As a first application, EAM potentials for decagonal Al-Ni-Co, 
icosahedral Ca-Cd, and both icosahedral and decagonal Mg-Zn  
quasicrystals have been constructed. The influence of the potential 
range and degree of specialisation on the accuracy and other properties 
is discussed and compared.

\vspace{.3cm}\noindent\emph{Keywords:} force matching; quasicrystal; effective
potential; molecular dynamics; ab initio
\end{abstract}

\section{Introduction}
\label{sec:intro}

Large-scale molecular dynamics simulations are possible only with classical
effective potentials, which reduce the quantum-mechanical interactions of
electrons and nuclei to an effective interaction between the atom cores.  The
computational task is thereby greatly simplified. Whereas ab-initio
simulations are limited to a few hundred atoms at most, classical simulations
can be done routinely with multi-million atom systems. For many purposes, such
system sizes are indispensable.  For example, fracture studies of
quasicrystals require samples with several million atoms at least
\cite{Roesch2005}. Diffusion studies, on the other hand, can be done with a
few thousand atoms (or even less), but require very large simulated times of
the order of nanoseconds \cite{Gahler2005}, which also makes them infeasible
for ab-initio simulations.

While physically justified effective potentials have been constructed for many 
elementary solids, such potentials are rare for complex intermetallic alloys.
For this reason, molecular dynamics simulations of these materials have often 
been done with simple model potentials, resulting in rather limited reliability
and predictability. In order to make progress, better potentials are needed
to accurately simulate complex materials. 

The force matching method \cite{Ercolessi1994} provides a way to construct
physically \red{reasonable} potentials also for more complex solids, where a
larger variety of local environments has to be described correctly, and many
potential parameters need to be determined. The idea is to compute forces and
energies from first principles for a suitable selection of small reference
systems, and to fit the potential parameters so that they optimally reproduce
these reference data. Hereafter, potentials generated in this way will be
referred to as fitted potentials. Thus, the force matching method allows to
make use of the results of ab-initio simulations also for large-scale
classical simulations, thereby bridging the gap between the sample sizes
supported by these two methods.

\section{Force Matching}
\label{sec:forcematching}

As we intend to construct potentials for complex intermetallic alloys, we have
to assume a functional form which is suitable for metals.  A good choice are
EAM (Embedded Atom Method) potentials \cite{Daw1984}, also known as
glue potentials \cite{Ercolessi1988}. Such potentials have been used
very successfully for many metals, and are still efficient to compute, even
though they include many-body terms. In contrast, pure pair potentials show a
number of deficiencies when it comes to describe metals \cite{Ercolessi1988}.  The functional form of
EAM potentials is given by
\begin{equation}
  \label{eq:gluepot}
  V=\sum_{i,j<i}\phi_{k_{i}k_{j}}(r_{ij})+
  \sum_{i}U_{k_{i}}
  (n_{i}),\qquad
\text{with}\qquad n_{i}=\sum_{j\neq i}\rho_{k_{j}}(r_{ij}),
\end{equation}
where $\phi_{k_{i}k_{j}}$ is a pair potential term depending on the two atom
types $k_{l}$. $U_{k_{i}}$ describes the embedding term that represents an
additional energy for each atom. This energy is a function of a local density
$n_{i}$ determined by contributions $\rho_{k_{j}}$ of the neighbouring atoms.
It is tempting to view this as embedding each atom into the electron sea
provided by its neighbours.  Such an interpretation is not really meaningful,
however.  The potential (\ref{eq:gluepot}) is invariant under a family of
``gauge'' transformations \cite{Ercolessi1988}, by which one can move
contributions from the embedding term to the pair term, and vice versa, so
that it makes little sense to give any of them an individual physical
interpretation.

In order to allow for maximal flexibility, and to avoid any bias, the 
potential functions in (\ref{eq:gluepot}) are represented by tabulated 
values and spline interpolation, the tabulated values acting as potential 
parameters. This makes it unnecessary to guess the right analytic form
beforehand. The sampling points can be chosen freely, which is useful 
for functions which vary rapidly in one region, but only slowly in another 
region.

The forces and energies in the reference structures are computed with VASP,
the Vienna Ab-Initio Simulation Package \cite{Kresse1993,%
  Kresse1996}, using the Projector Augmented Wave (PAW) method
\cite{Blochl1994,Kresse1999}.  Like all plane wave based ab-initio codes, VASP
requires periodic boundary conditions. For quasicrystals, this means that
periodic approximants have to be used as reference structures. As ab-initio
methods are limited to a few hundred atoms, those approximants must be rather
small. For the systems studied so far, this was not a major problem, as the
relevant local environments in the quasicrystal all occur also in reasonably
small approximants. Icosahedral quasicrystals with F-type lattice may be more
problematic in this respect. For these, small approximants are rare, and the
force matching method requires a sufficient variety of reference structures.

Given the reference data (forces, energies, and stresses in the reference
structures), the potential parameters 
(in our case: up to
about 120 EAM potential sampling points for spline interpolation)
 then are
optimised in a non-linear least square fit, so that the fitted potential
reproduces the reference data as well as possible. The target function to be
minimised is a weighted sum of the squared deviations between the reference
data, denoted by the subscript 0 below, and the corresponding data computed
from the fitted effective potential. It is of the form
\begin{eqnarray}
  \label{eq:target}
  Z&=&Z_{\text{F}}+Z_{\text{C}},\qquad \text{with}\\
Z_{\text{F}}&=&\sum\limits_{j=1}^{N_{A}}\sum\limits_{\alpha=x,y,z}W_{j}
\frac{\left(f_{j_{\alpha}}-f_{0,j_{\alpha}}\right)^{2}}
{\boldsymbol{f}_{0,j}^{2}+\varepsilon_{j}},\quad \text{and} \quad
Z_{\text{C}}=\sum\limits_{k=1}^{N_{c}}W_{k}
\frac{\left(A_{k}-A_{0,k}\right)^{2}}{A_{0,k}^{2}+\varepsilon_{k}},\label{eq:tar2}
\end{eqnarray}
where $Z_{F}$ represents the contributions of the forces $\boldsymbol{f}_{j}$,
and $Z_{C}$ those of some collective quantities like total stresses and
energies, but also additional constraints $A_{k}$ on the potential
one would like to impose.
The denominators of the fractions ensure that the target function measures the
relative deviations from the reference data, except for really tiny
quantities, where the $\varepsilon_{l}$ prevent extremely small denominators.
The $W_{l}$ are the weights of the different terms. It proves useful for the
fitting to give the total stresses and the cohesion energies a higher weight,
although in principle they should be reproduced correctly already from the
forces.

We developed a programme named \emph{potfit}, which optimises the potential
parameters to a set of reference data. It consists of two largely independent
parts. The first part implements a particular parametrised potential model. It
takes a list of potential parameters and computes from it the target function,
i.e., the deviations of the forces, energies, and stresses from the reference
data. Wrapped around this part is a second, potential independent part, which
implements a least square minimisation module, using a combination of a
deterministic conjugate gradient algorithm \cite{Powell1965} and a
stochastic simulated annealing algorithm \cite{Corana1987}. This part
knows nothing about the details of the potential, and only deals with a list
of potential parameters. The programme architecture thus makes it easy to
replace the potential dependent part by a different one, e.g., one which
implements a different potential model, or a different way to parametrise it.

\section{Results and Applications}

\begin{figure}[tb]
\centerline{\includegraphics[width=\textwidth]{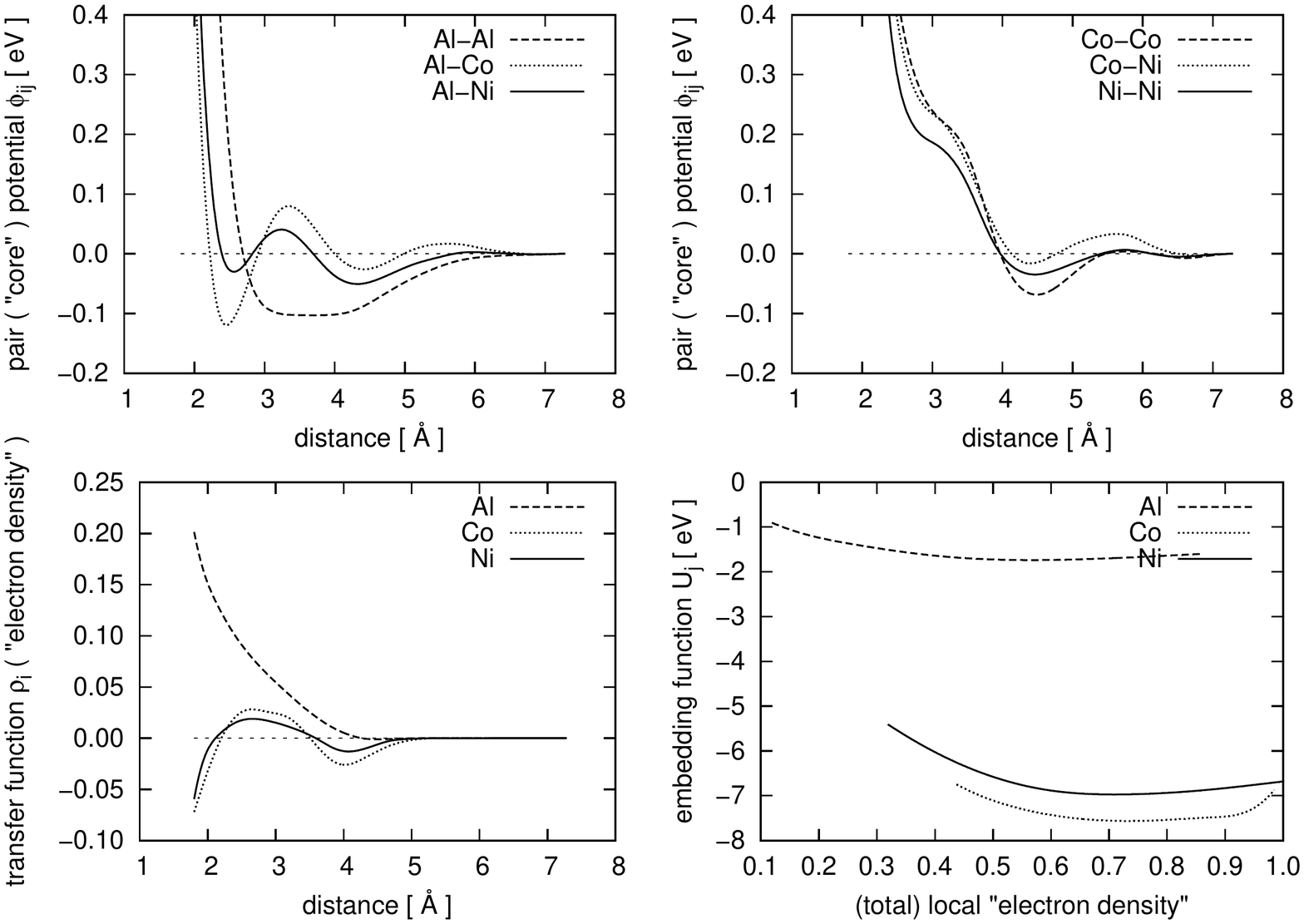}}
\caption{Potential functions for decagonal Al-Ni-Co}
\label{AlNiCo-pot}
\end{figure}

We generated several fitted potentials for decagonal Al-Ni-Co and
icosahedral Ca-Cd quasicrystals, as well as Mg-Zn potentials suitable for both
icosahedral and decagonal phases. In a first step, classical molecular
dynamics simulations with simple model potentials were used to create
reference configurations from small approximants (80--250 atoms). These
included samples at different temperatures, but also samples which were scaled
and strained in different ways. The approximants were carefully selected, so
that all relevant local environments are represented. For those reference
structures, the forces, stresses and energies were computed with ab-initio
methods, and a first version of the 
fitted effective potential
given by sampling points with cubic spline interpolation
was fitted to the reference data. In a
second step, molecular dynamics simulations with the newly determined
potential were used to create new reference structures, which are better
representatives of the structures actually appearing in that system. The new
reference structures complemented and partially replaced the previous ones,
and the fitting procedure was repeated. This second iteration resulted in a
significantly better fit to the reference data. In order to test the
transferability of the 
fitted potentials, 
further samples similar to the reference
structures were created, and their ab-initio forces and energies were compared
to those determined by the classical potentials. The deviations were of the
same order as the deviations found in the potential fit, which shows that the
fitted potentials transfer well to similar structures.
For Al-Ni-Co, a
force-matched potential is displayed in figure \ref{AlNiCo-pot}.
Fitted potentials for Ca-Cd and Mg-Zn are not displayed here for
  space constraints, but are available from the authors.

The potentials developed for decagonal Al-Ni-Co quasicrystals are intended to
be used in high-temperature diffusion simulations \cite{Gahler2005}.  It is
therefore important that they describe high temperature states well, which is
achieved by selecting the reference structures accordingly.  By using high
temperature reference structures, 
the fitted potential is especially
trained to such situations. As part of the potential validation, the melting
temperature was determined by slowly heating the sample at constant pressure,
and the elastic constants of decagonal Al-Ni-Co were determined.  We actually
have constructed two potential variants: Variant A gives excellent values for
the elastic constants (Table \ref{tab:elast}), but produces a melting
temperature which is somewhat too high. Conversely, variant B shows larger
deviations in the elastic constants, but gives a very reasonable value of the
melting temperature of about 1300~K.  It is a general experience that with an
effective potential it is often not possible to reproduce all desired
quantities equally well at the same time.

\begin{table}[tb]
\begin{center}
\caption{Elastic constants of decagonal Al-Ni-Co\label{tab:elast}}
\medskip
\begin{tabular}[t]{l|cccccc}
[GPa] & $c_{11}$ & $c_{33}$ &  $c_{44}$ & $c_{66}\,^{a}$ & $c_{12}$ & 
 $c_{13}$ \\\hline
Exp. \cite{Chernikov1998} & 234 & 232 & 70 & 88 & 57 & 67 \\
Pot. A                  & 230 & 231 & 55 & 70 & 91 & 91 \\
Pot. B                  & 197 & 187 & 49 & 58 & 86 & 84 \\
 \multicolumn{7}{l}{$^{a}$\footnotesize 
In decagonal QC: $c_{66}=\frac{1}{2}(c_{11}-c_{12})$ }
\end{tabular}
\end{center}
\end{table}

In complex intermetallic systems there are many competing candidates for the
ground state structure. 
This is the case also for complex crystalline
systems. In principle, the ground state of these can be determined 
directly by ab-initio simulations, but for large unit cells this is 
extremely time-consuming, or even impossible. 
Classical potentials can be used to select
the most promising candidates, and to pre-relax them, so that the time for
ab-initio relaxation can be dramatically reduced.  Potentials used for this
purpose must be able to discriminate energy differences of the order of a
meV/atom. This has been largely achieved with 
fitted potentials for the Mg-Zn and
Ca-Cd systems, by using mainly near ground state structures as reference
structures. Also, for this application it is important to choose a small
$\varepsilon_{j}$ in equation (\ref{eq:tar2}), so that small forces are also
reproduced accurately. 
The so constructed Ca-Cd potentials
  have been used successfully for structure optimisations \cite{Widom2005}.

\section{Discussion and Conclusion}

The selection of the reference structures used for the potential fit
largely determines the capabilities of the 
resulting potential. For a precise
determination of the ground state, low temperature structures 
should be dominant in the reference structures, and it must be assured 
that even small forces and energy differences are reproduced accurately. 
For high temperature simulations, on the other hand, typical high 
temperature structures must be predominant in the reference structures.
This opens the possibility to design specialised potentials for
certain purposes by a suitable selection of reference structures.
It should be kept in mind, however, that a 
fitted
potential can only deal 
with situations it has been trained to. For instance, one should 
not expect a 
fitted
potential to handle surfaces correctly, if it was trained 
only with bulk systems. Clearly, there is always a trade-off between the 
transferability and the accuracy of a 
fitted potential.
A potential can be 
made more versatile by training it with many different kinds of structures, 
but the more versatile it becomes, the less accurate it will be on average.
Conversely, very accurate 
fitted potentials 
will probably have limited
transferability. 

For practical applications, the range of a potential is also an important
issue, as it enters in the third power in the computational effort of 
molecular dynamics. Allowing for a larger potential range results in 
greater flexibility of the potential, which might improve its accuracy,
but this comes at the price of a slower simulation. We therefore need
a compromise between speed and accuracy. The potential range should only
be increased as long as this can improve the potential quality. In a
first step, our 
fitted potentials 
were constructed with a fairly generous 
range of about 7\AA. It turned out, however, that especially the
transfer function $\rho_{i}$ did not make effective use of this range,
and was essentially zero beyond 5\AA. In a second fit we therefore
restricted the range of $\rho_{i}$ to 5\AA, without significant loss
of accuracy. This is one of the advantages of using tabulated functions:
The system itself chooses the optimal functions, including the optimal
range.

Force Matching has proven to be a versatile method to construct physically
reasonable, 
accurate effective potentials even for structures as complicated as
quasicrystals and their approximants. Our \emph{potfit} programme makes it easy
to apply this method to different systems, and is also easy to adapt for the
support of further potential models. The potentials constructed so far have
successfully been used in high temperature diffusion simulations of decagonal
Al-Ni-Co \cite{Gahler2005}, and in structure optimisation of
approximants in the Zn-Mg and Ca-Cd systems. Further fruitful applications of
the 
fitted potentials 
can certainly be expected, and we hope to apply our methods
also to other complex alloy systems, where reliable potentials are still
lacking.

\section*{Acknowledgement}
This work was funded by the Deutsche Forschungsgemeinschaft through
Sonderforschungsbereich 382. Special thanks go to Marek Mihalkovi\v{c} for
supplying approximants and feedback in the Ca-Cd and Mg-Zn systems, and to
Hans-Rainer Trebin for supervising the thesis work of the first author.

\section*{References}
\vspace{-.5cm}

\begin{thebibliography}{10}

\bibitem{Roesch2005}
F.~{R{\"o}sch}, Ch.~{Rudhart}, J.~{Roth}, H.-R.~{Trebin}, and P.~{Gumbsch},
\emph{Phys.\ Rev.~B} \textbf{72}, 014128 (2005).

\bibitem{Gahler2005} S.~{Hocker}, F.~{G\"ahler}, and P.~{Brommer},
  \emph{Phil.\ Mag.} \textbf{86}, 1051 (2006).

\bibitem{Ercolessi1994}
F.~{Ercolessi} and J.~B. {Adams}, \emph{Europhys.\ Lett.} \textbf{26}, 583
  (1994).

\bibitem{Daw1984}
M.~S. {Daw} and M.~I. {Baskes}, \emph{Phys.\ Rev.~B} \textbf{29}, 6443 (1984).

\bibitem{Ercolessi1988}
F.~{Ercolessi}, M.~{Parrinello}, and E.~{Tosatti}, \emph{Phil.\ Mag.~A}
  \textbf{58}, 213 (1988).

\bibitem{Kresse1993}
G.~{Kresse} and J.~{Hafner}, \emph{Phys.\ Rev.~B} \textbf{47}, 558 (1993).

\bibitem{Kresse1996}
G.~{Kresse} and J.~{Furthm{\"u}ller}, \emph{Phys.\ Rev.~B} \textbf{54}, 11169
  (1996).

\bibitem{Blochl1994}
P.~E. {Bl{\"o}chl}, \emph{Phys.\ Rev.~B} \textbf{50}, 17953 (1994).

\bibitem{Kresse1999}
G.~{Kresse} and D.~{Joubert}, \emph{Phys.\ Rev.~B} \textbf{59}, 1758 (1999).

\bibitem{Powell1965}
M.~J.~D. {Powell}, \emph{Comp.\ J.} \textbf{7}, 303 (1965).

\bibitem{Corana1987}
A.~{Corana}, M.~{Marchesi}, C.~Martini, and S.~Ridella, \emph{ACM Trans.\
  Math.\ Soft.} \textbf{13}, 262 (1987).


\bibitem{Chernikov1998}
M.~A. {Chernikov}, H.~R. {Ott},  A. Bianchi, A.~Migliori, and T.~W.~Darling, \emph{Phys.\ Rev.\ Lett.} \textbf{80}, 321 (1998).

\bibitem{Widom2005}
  M.~Mihalkovi{\v{c}} and M.~{Widom}, \emph{Phil.\ Mag.} \textbf{86}, 519
  (2006). 

\end{thebibliography}

\end{document}